\documentclass[aps,prb,floatfix,twocolumn,showpacs,nofitinbib,superscriptaddress,hyperref=pdftex,citeautoscript]{revtex4-1}
\usepackage{epsfig}
\usepackage{amsmath}
\usepackage{amssymb}
\usepackage{amsfonts}
\usepackage{ dsfont }
\usepackage{ comment,color }
\usepackage{lipsum}
\usepackage{hyperref,mathtools}
\usepackage{makecell}
\usepackage{array}
\usepackage[sep=3pt, offset=1em]{simpler-wick}

\hypersetup{colorlinks=true,linkcolor=blue,anchorcolor=blue,citecolor=blue,filecolor=blue,urlcolor=blue,bookmarksnumbered=true,pdfview=FitB}

\newcommand{\bk}{{{\boldsymbol{k}}}}

\newcommand{\bQ}{{{\boldsymbol{Q}}}}
\newcommand{\br}{{{\boldsymbol{r}}}}
\newcommand{\bd}{{{\boldsymbol{d}}}}

\newcommand{\bq}{{\boldsymbol{q}}}

\newcommand{\bkappa}{{\boldsymbol \kappa}}
\newcommand{\bp}{{\boldsymbol{p}}}
\newcommand{\bb}{{\boldsymbol{b}}}

\newcommand{\beqa}{\begin{eqnarray}}
\newcommand{\eeqa}{\end{eqnarray}}

\newcommand{\ua}{\uparrow}
\newcommand{\da}{\downarrow}

\newcommand{\bo}{{\boldsymbol{0}}}

\newcommand{\bdelta}{{\boldsymbol \delta}}

\begin{document}

\hsize\textwidth\columnwidth\hsize\csname@twocolumnfalse\endcsname

\title{Nematic, chiral and topological superconductivity in transition metal dichalcogenides}

\author{Constantin Schrade}
\affiliation{Hearne Institute of Theoretical Physics, Department of Physics \& Astronomy, Louisiana State University, Baton Rouge LA 70803, USA}

\author{Liang Fu}
\affiliation{Department of Physics, Massachusetts Institute of Technology, 77 Massachusetts Ave., Cambridge, MA 02139}
\date{\today}

\vskip1.5truecm
\begin{abstract}
We introduce and study a realistic model for superconductivity in twisted bilayer WSe$_{2}$, where electron pairing arises from spin-valley fluctuations in the weak-coupling regime. Our model comprises both the full continuum model moir\'{e} bandstructure and a short-ranged repulsive interaction. By calculating the spin-valley susceptibility, we identify a Fermi surface nesting feature near half-filling of the top-most moir\'{e} band, which induces significantly enhanced spin-valley fluctuations. We then analyze the dominant Kohn-Luttinger pairing instabilities due to these spin-valley fluctuations and show that the leading instability corresponds to a two-component order parameter, which can give rise to nematic, chiral and topological superconductivity. As our findings are asymptotically exact for small interaction strengths, they provide a viable starting point for future studies of superconductivity in twisted transition metal dichalcogenide bilayers. 
\end{abstract}


\maketitle
\section{Introduction}
Two-dimensional transition metal dichalcogenides (TMDs) have in recent years emerged as a promising material platform for realizing a plethora of new electronic phases \cite{bib:Manzeli2017}. Prominent examples of these electronic phases in monolayer TMDs include the quantum spin Hall effect \cite{bib:Qian2014,bib:Tang2017,bib:SWu2018,bib:Shi2019}, arising due to the strongly spin-orbit coupled band structure, and Ising superconductivity \cite{bib:Lu2015,bib:Xi2016,bib:Sohn2018,bib:Sergio2018}, which is realized as a result of the substantial effective electron mass that enhances the importance of interaction effects.  

Beyond the TMD monolayers, another research frontier that is currently evolving at rapid pace are moir\'{e} lattices realized with multilayer TMDs \cite{bib:LeRoy2020,bib:Regan2020,bib:Yang2020,bib:Jin2021,bib:Ghiotto2021,bib:Wang2020,bib:Li2021,bib:Li2021_2,bib:Millis2021}. In these systems, a lattice mismatch or a rotational misalignment generates an effective superlattice that can be utilized for simulating strongly-correlated electron states \cite{bib:Wu2018,bib:Wu2019,bib:Schrade2019,bib:Pan2020,bib:Pan2020_2,bib:Zhang2020,bib:Zhang2020_2,bib:Devakul2021,bib:Zhang2021,bib:Hsu2021}. For example, in WSe$_{2}$/WS$_{2}$ heterobilayers, experiments have identified Mott insulators \cite{bib:Regan2020,bib:Yang2020}, Wigner crystals \cite{bib:Regan2020}, and stripe-ordered states \cite{bib:Jin2021}. Moreover, in twisted WSe$_{2}$ (tWSe$_{2}$) homobilayers, experiments have found evidence for quantum criticality \cite{bib:Ghiotto2021} and correlated insulating states \cite{bib:Wang2020}. Notably, the discovered correlated insulators in tWSe$_{2}$ appeared at half-filling for broad twist angle range, $\theta\sim4^{\circ}$-$5.1^{\circ}$, for which the bandwidth is likely comparable to the interaction strength. This weak- to intermediate coupling scenario is further supported by a resistivity enhancement upon tuning a van-Hove singularity to half-filling via an external displacement field. Both features highlight the \textit{importance of bandstructure effects} for understanding correlated electron states in tWSe$_{2}$ \cite{bib:Bi2021}.

In addition to the discovery of correlated insulators, an experiment on tWSe$_{2}$ also found a ``zero-resistance state'' when doping away from half-filling at $\theta\sim5.1^{\circ}$ \cite{bib:Wang2020}. The appearance of this zero-resistance state indicates that, similar to graphene-based superlattices \cite{bib:Cao20182,bib:Yankowitz2019,bib:Chen20192,bib:Liu2020,bib:Park2021,bib:Hao2021,bib:Cao2021}, tWSe$_{2}$ can also host superconductivity. However, it is so far unclear what properties of the bandstructure can effect the candidate superconducting state in tWSe$_{2}$. Moreover, it is also an open question on  what constitutes the pairing mechanism and pairing symmetry. 

 \begin{figure}[!t] \centering
\includegraphics[width=\linewidth] {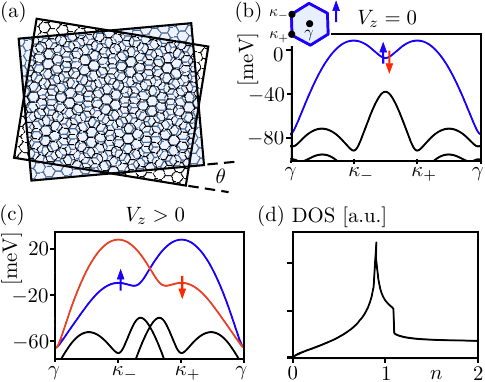}
\caption{(Color online)
(a) WSe$_{2}$ layers at twist angle $\theta$ forming a triangular moir\'{e} lattice. 
(b) Bandstructure along high-symmetry lines for $(\theta,V_{z},V,\psi,w)=(5.1^{\circ},0\,\text{meV},9\,\text{meV},128^{\circ},18\,\text{meV})$. Inset shows the moir\'{e} Brillouin zone at the $+K$ valleys. (c) Same as (b) but for $V_{z}=42\,\text{meV}$. The spin-up (blue) and spin-down (red) bands at $\pm K$ valleys
are split by the finite displacement field. (d) DOS with a maximum near $n_{\text{max}}\approx0.9$.
}\label{fig:1}
\end{figure}

Here, we address these questions by introducing a model for superconductivity in tWSe$_{2}$, where electron pairing arises from spin-valley fluctuations. Our model comprises both the moir\'{e} bandstructure of tWSe$_{2}$ and a short-ranged repulsive interaction. 
Near half-filling of the top-most moir\'{e} band, we demonstrate that the Fermi surfaces of the $\pm K$ valleys exhibit a strong nesting feature. This nesting maps the occupied states of one valley onto the other valley's unoccupied states, leading to strong spin-valley fluctuations especially near the insulating state \cite{bib:Bi2021}. We show that, interestingly, electron pairing mediated by such enhanced spin-valley fluctuations can give rise to nematic, chiral and topological superconductivity. Our findings are asymptotically exact in the weak coupling limit \cite{bib:Raghu2010} and provide a starting point for studies of superconductivity in TMD systems. 

\section{Model}
\subsection{Continuum model Hamiltonian}
We consider two layers of WSe$_{2}$ that are initially aligned in the AA-stacking configuration and, subsequently, rotated by a small twist angle $\theta$, see Fig.\,\ref{fig:1}(a). In this situation, the $\pm K$ valleys of the layers have comparable energy so that inter-layer tunneling becomes effective, which leads to a layer-hybridized moir\'{e} bandstructure. For the discussion of this moir\'{e} bandstructure, we focus separately on the spin-up (spin-down) valence bands of the $+K$ ($-K$) valleys. This approach is justified by the substantial valley-dependent spin splitting in the valence bands and the effective decoupling of the valleys due to their large momentum space separation. 
The Hamiltonian for the continuum model moir\'{e} bands of spin-up electrons at the $+K$ valleys then reads \cite{bib:Wu2019}, 
 \begin{equation}
 H_{\uparrow}=
 \int
 \mathrm{d}\br\,
 \Psi_{\ua}^{\dag}(\br)
 \begin{pmatrix}
   h^{t}_{\bk\ua}(\br) & \Delta_{T}(\br) \\
     \Delta^{\dag}_{T}(\br) &  h^{b}_{\bk\ua}(\br)
 \end{pmatrix}
  \Psi_{\ua}(\br),
 \label{Eq1}
 \end{equation}
 where $\Psi_{\ua}(\br)=(c^{t}_{\ua}(\br),c^{b}_{\ua}(\br))^{T}$ is an electron spinor with components corresponding to the top/bottom $(t/b)$ layer.
 The Hamiltonian for spin-down electrons at the $-K$ valleys is obtained from $H_{\uparrow}$ via a time-reversal operation.
 
 First, we discuss the diagonal components of $H_{\uparrow}$, which represent the individual layers. They are given by, 
 \begin{equation}
h^{t/b}_{\bk\ua}(\br)=-\frac{\hbar^{2}(\bk-\bkappa_{\pm})^{2}}{2m^{*}}\pm \frac{V_{z}}{2}+\sum_{j=1,3,5}2V\cos(\bb_{j}\cdot\br\pm\psi).
\end{equation}
Here, the first term describes the kinetic energy with effective mass $m^{*}=0.43m_{0}$ ($m_0$ is the bare electron mass) and momentum shifts $\bkappa_{\pm}=\left[4\pi|\theta|/(3a_{0})\right](-\sqrt{3}/2,\mp1/2)$ ($a_{0}=3.317\,\text{\AA}$ is the monolayer lattice constant), that account for for the layer rotation in momentum space. The second term corresponds to a layer potential difference $V_{z}$ due to an out-of-plane displacement field. The third term models the moir\'{e} potential with amplitude $V$, phase offset $\psi$, and reciprocal lattice vectors $\bb_{j}=C^{j-1}_{3}(4\pi|\theta|/\sqrt{3}a_{0},0)$ ($C_{3}$ is a $2\pi/3$ rotation). 

Besides the individual layer terms, $H_{\uparrow}$ also includes off-diagonal terms for inter-layer coupling. They read, 
\begin{equation}
\Delta_{\text{T}}(\br)=w(1+e^{-i\bb_{2}\cdot\br}+e^{-i\bb_{3}\cdot\br}), 
\end{equation}
where $w$ corresponds to the inter-layer coupling strength. 

Having defined the continuum model Hamiltonian, we now proceed by analyzing its spatial symmetries. We find that the continuum model exhibits a three-fold rotation symmetry given by $U^{\dag}_{C_3}(\br)H_{C_{3}(\bk)\ua }(C_{3}(\br))U_{C_3}(\br)=H_{\bk\ua}(\br)$ with the Hamiltonian density $H_{\bk\ua}(\br)$, $U_{C_3}(\br)=e^{i(1+\tau_{z}/2)\bb_{2}\cdot\br_{2}}e^{i\bb_{3}\cdot\br}$, and a layer-space Pauli matrix $\tau_{z}$. Besides the rotation symmetry, which is also a microscopic symmetry of the moir\'{e} lattice, the continuum model exhibits an emergent mirror symmetry along $y=0$ given by $H_{M_x(\bk)\ua }(M_x(\br))=H_{\bk\da}(\br)$. Interestingly, this mirror symmetry flips the two valleys.

\subsection{Moir\'{e} bandstructure}
 As a final step, we plot the moir\'{e} bands along high-symmetry lines of the Brillouin zone for $(\theta, V,\psi,w)=(5.1^{\circ},9\,\text{meV},128^{\circ},18\,\text{meV})$ \cite{bib:Devakul2021} with $V_{z}=0\,\text{meV}$ and $V_{z}=42\,\text{meV}$, see Fig.\,\ref{fig:1}(b) and (c). We find that finite displacement field induces an approximate saddle point in the top-most moir\'{e} bands $\xi_{\ua/\da}(\bk)$ at $\bkappa_{\pm}$. At these points, the Fermi velocity is greatly reduced, which leads to an enhancement of the density of states (DOS) for a filling of $n\approx0.9$ electrons per moir\'{e} unit cell, see Fig.\,\ref{fig:1}(d). The emergence of an enhanced DOS near $n\approx1$ is in accordance with experiments \cite{bib:Wang2020} and signals that correlation effects due to spin-valley fluctuations may be amplified. In our further analysis, we will fix above choice for $(\theta, V,\psi,w)$ that we obtained from density functional calculations \cite{bib:Devakul2021} and the field value $V_{z}=42\,\text{meV}$. 

\section{Spin-valley fluctuations}
We now discuss the ordering instabilities that arise from the spin-valley fluctuations. To identify the spin-valley fluctuations with dominant modulation vectors, we will analyze the maxima of the spin-valley susceptibility,  
\begin{equation}
\chi_{\alpha\beta}(\bq,\omega)
=
-
\frac{1}{\beta N}
\sum_{\bk}
\text{Tr}\left[\sigma_{\alpha}G(\bk,i\omega)
\sigma_{\beta}G(\bk+\bq,i\omega)\right].
\end{equation}
In this definition, $\sigma_{\alpha,\beta}\in\{\sigma_{z},\sigma_{\pm}=(\sigma_{x}\pm i\sigma_{y})/2\}$ are spin/valley-space Pauli matrices, `Tr' denotes the trace over the spin/valley indices, $\beta$ is the inverse temperature, and $N$ is the number of superlattice unit cells. Furthermore, the non-interacting Green's function is given by, 
\begin{equation}
G_{s_{1}s_{2}}(\bk,i\omega)
=
\delta_{s_{1}s_{2}}
[
i\omega-\xi_{s_{1}}(\bk)
+\mu
]^{-1}.
\end{equation}
Here, $\xi_{s}(\bk)$ are the dispersions for the top-most moir\'{e} band of the $\pm K$ valleys with $\xi_{\da}(\bk)=\xi_{\ua}(-\bk)$ due to time-reversal symmetry, $\mu$ is the chemical potential, and $\omega$ corresponds to a fermionic Matsubara frequency.

 \begin{figure}[!t] \centering
\includegraphics[width=\linewidth] {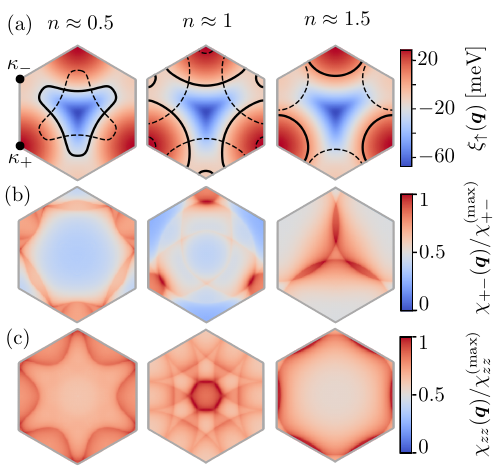}
\caption{(Color online)
(a) Dispersion $\xi_{\ua}(\bq)$ of the top-most band at the $+K$ valleys. The  $+K$ ($-K$) valley Fermi surfaces are shown with solid (dashed) black lines. (b) Transversal spin-valley susceptibility $\chi_{+-}(\bq)$ at $1/\beta=100\,$meV. 
(c) Longitudinal spin-valley susceptibility $\chi_{zz}(\bq)$ at $1/\beta=100\,$meV. }\label{fig:2}
\end{figure}

Having defined the spin-valley susceptibility, we will now investigate its static longitudinal and transversal components, $\chi_{zz}(\bq)=\chi_{zz}(\bq,0)$ and $\chi_{+-}(\bq)=\chi_{+-}(\bq,0)$. In our model, these two components of the spin-valley susceptibility will show different dependencies on the modulation vector $\bq$ as the external displacement
fields lifts the $SU(2)$ valley rotation symmetry \cite{bib:Pan2020}.

First, we focus on the filling of $n\approx1$ electron per moir\'{e} unit cell because it is the experimentally most relevant case. In this situation, the Fermi surfaces have the shape of triangles centered around the $\gamma$ point of the Brillouin zone as shown in Fig.\,\ref{fig:2}(a). 
As shown in Fig.\,\ref{fig:2}(b), we numerically find that $\chi_{+-}(\bq)$ is enhanced near $\bq\approx\bQ_{j}$ with $\bQ_{j}=C^{j-1}_{3}\bkappa_{+}$, indicating an instability towards an inter-valley excitonic insulator that pairs electrons and holes of opposite valleys \cite{bib:Bi2021}.
However, due to the broken $SU(2)$ valley rotation symmetry, the same features are \textit{not} present in the longitudinal valley susceptibility. Instead, as shown in Fig.\,\ref{fig:2}(c), $\chi_{zz}(\bq)$ is enhanced near $\bq\approx \bo$, suggesting that the excitonic instability competes with a spin/valley-polarized state \cite{bib:Devakul2021}. 

We now compare these results with the situation away from $n\approx1$. For fillings $n\lesssim0.9$, the Fermi surfaces shrink to smaller triangles and  the dominant modulation vectors in the spin-valley susceptibility are modified. For example, at $n\approx0.5$, $\chi_{+-}(\bq)$ is enhanced near $\bq\approx\bQ'_{j}$ with $\bQ'_{j}=C^{j-1}_{3}\bkappa_{-}$, while $\chi_{zz}(\bq)$ shows maxima near the Brillouin zone boundaries. For $n>1$, the Fermi surfaces are center at the Brillouin zone corners. At $n\approx1.5$, $\chi_{+-}(\bq)$ is enhanced near $\bq\approx \bo$, while $\chi_{zz}(\bq)$ is again enhanced near the Brillouin zone edges. 

\section{Superconductivity}
\subsection{Kohn-Luttinger mechanism}
We will now demonstrate that spin-valley fluctuations can mediate superconductivity by a Kohn-Luttinger mechanism that realizes Cooper pairing from a nominally repulsive interaction term in the weak-coupling regime\cite{bib:Kohn1965,bib:Raghu2010,bib:Chubukov1993,bib:Maiti2013}. 

Before discussing the details of our approach, we highlight that the Kohn-Luttinger mechanism has also been valuable for understanding unconventional superconductivity in the intermediate-coupling regime and has already been applied to graphene systems\cite{bib:Lin2018,bib:Gonzalez2019,bib:Ghazaryan2021,bib:You2022,bib:Guinea2022}. For $t$WSe$_{2}$, bandstructure effects remain critical for understanding correlated states in these samples. In particular, by tuning the van-Hove singularity in the moiré bandstructure close to half-filling, a significant increase in the longitudinal resistance peak was noted, which indicates the emergence of the correlated insulator state\cite{bib:Wang2020}. Therefore, we expect that the Kohn-Luttinger mechanism can offer qualitative insights into the realized superconducting pairing symmetry.

To start, we introduce the Hubbard Hamiltonian, 
\begin{equation}
H=\sum_{\bk s}\xi_{s}(\bk) c^{\dag}_{\bk s}c_{\bk s}+\frac{U}{N}\sum_{\substack{\bk,\bp,\bq \\ s, s'}} c^{\dag}_{\bq s}c^{\dag}_{\bp+\bk-\bq s'}c_{\bp s'}c_{\bk s},
\end{equation}
where the quadratic part comprises the dispersion $\xi_{s}(\bk)$ of the top-most moir\'{e} band of the continuum model and $c_{\bk s}$ is the electron annihilation operator with momentum $\bk$ and spin $s$. In particular, as we omit multi-band effects, there is an emergent mirror symmetry $M_{y}$ along $x=0$ with $\xi_{s}(M_{y}\bk)=\xi_{s}(\bk)$. The interaction term in the Hamiltonian is taken as a local (compared to moir\'e scale) repulsive interaction with a strength $U>0$.

Next, we compute the Cooper channel interaction vertex, which describes scattering of electrons with spin polarization $s$ and $s'$ from the Fermi surface momenta $(\hat\bk',-\hat\bk')$
to $(\hat\bk,-\hat\bk)$. In general, the interaction vertex comprises a contribution from opposite-spin and equal-spin scattering. However, the equal-spin scattering leads to pairing with a `$\bd$-vector'
pointed orthogonal to the spin polarization \cite{bib:Sigrist2005}. Such an equal-spin pairing exhibits a suppressed critical temperature \cite{bib:Frigeri2004,bib:Frigeri2004_2} and, hence, our focus will be on opposite-spin scattering. To second order in $U$, we find that the interaction vertex \cite{bib:supplemental}, 
\begin{equation}
\label{Eq7}
\Gamma(\hat\bk,\hat\bk')=U+U^{2}\chi_{+-}(\hat\bk+\hat\bk').
\end{equation}
Notably, this result is different for TMD monolayers, where a parabolic dispersion induces a momentum dependence only at order $U^{3}$ \cite{bib:Hsu2014}. 

Next, we perform a mean-field decoupling of the effective pairing interaction and compute
the superconducting order parameter from the self-consistency equation \cite{bib:supplemental}, 
\begin{equation}
\label{Eq8}
\int_{\text{FS}_{\ua}} \frac{d\hat\bk'}{v_{\ua}(\hat\bk')}
\Gamma(\hat\bk,\hat\bk')
\Delta(\hat\bk')
=
\lambda
\Delta(\hat\bk), 
\end{equation}
Here, $v_{\ua}(\bk)=||\partial \xi_{\ua}(\bk)/\partial\bk||$ is the Fermi velocity and $\Delta(\hat\bk)$ is the superconducting
order parameter along the $+K$ valley Fermi surface, $\xi_{\ua}(\bk)=\mu0$. The order parameters at the $-K$ Fermi surface, $\xi_{\da}(\bk)=\mu$, is $\Delta'(\hat\bk)=-\Delta(-\hat\bk)$. Three remarks about Eq.\,\eqref{Eq8} are in order: 

 \begin{figure}[!t] \centering
\includegraphics[width=\linewidth] {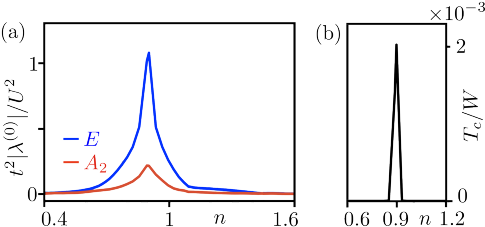}
\caption{(Color online)
(a) Pairing strength given by the most negative eigenvalue $\lambda^{(0)}$ of the the self-consistency 
equation as a function of filling $n$ with $t=\hbar^{2}\theta^{2}/2m^{*}a^{2}_{0}$. For fillings $1.4 \lesssim n \lesssim 1.45$,
the most negative eigenvalue is in the $1d$ $A_{2}$-representation of the $C_{3v}$ point group. At all other fillings, the most negative eigenvalue is in the $2d$ $E$-representation. (b) Critical temperature normalized by the bandwidth $W$ for $(U/t)^{2}=0.15$, showing a pronounced enhancement around $n\approx n_{\text{max}}=0.9$. Taking a bandwidth $W\approx80\,$meV (see Fig.\,1(a)), we have $T_{c}\approx 1.85\,$K.
}\label{fig:3}
\end{figure}

(1) The negative eigenvalues $\lambda<0$  correspond to sectors with an attractive interaction in the Cooper channel and grow in magnitude under renormalization. In particular, the most negative eigenvalue $\lambda^{(0)}$ is most relevant in the renormalization group sense, triggering the leading superconducting instability at $T_{c}\propto \exp(-1/|\lambda^{(0)}|)$ \cite{bib:Raghu2010}. 

(2) The eigenvectors $\Delta(\bk)$ correspond to form factors of the superconducting order parameters and transform as irreducible representations $\{A_{1},A_{2},E\}$ of the  point group $C_{3v}$ that is generated by $C_{3}$ and $M_y$. In particular, since $C_{3v}$ has no inversion center, singlet and triplet pairings mix in tWSe$_{2}$ \cite{bib:Hsu2014,bib:Yuan2014}.  

(3) In the weak coupling limit, the term $\propto U$ in $\Gamma(\hat\bk,\hat\bk')$ is parametrically larger than the term $\propto U^{2}$. For an attractive interaction in the Cooper channel, the eigenvectors $\Delta(\bk)$ thus need to annihilate the first-order term $\propto U$ exactly in the expression for interaction vertex of Eq.\,\eqref{Eq7} \cite{bib:Raghu2010,bib:Cho2013}. Or, explained differently, the eigenvectors $\Delta(\bk)$ need to be in the null-space of the matrix representation of the first-order term $\propto U$ in Eq.\,\eqref{Eq7}. 
Such an annihilation implies that $s$-wave pairing, $\Delta^{(s)}(\bk)\propto 1$, is suppressed. In particular, the $s$-wave pairing with $\Delta(\boldsymbol{k},\boldsymbol{k'}) \propto 1$ cannot annihilate this first-order contribution and thus cannot be realized. Furthermore, the absence of inversion symmetry in our system results in a mixture of singlet and triplet pairing channels. Consequently, suppression of $s$-wave pairing extends to other channels within the $A_{1}$ representation\cite{bib:Hsu2014}. Hence, the pairing needs to arise from the remaining representations, which are the $1d$ $A_{2}$ representation or the $2d$ $E$-representation.

 \begin{figure}[!t] \centering
\includegraphics[width=\linewidth] {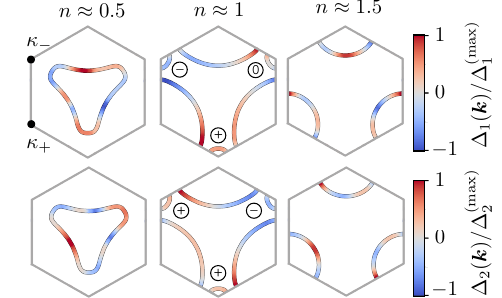}
\caption{(Color online)
Linear independent gap functions of the leading superconducting state in the $E$ representation. 
}\label{fig:4}
\end{figure}

\subsection{Superconducting pairings}
To determine the leading eigenvalue $\lambda^{(0)}$ and the representation of the associated superconducting order parameter, we solve Eq.\,\eqref{Eq8} numerically for a broad range of fillings $n\in [0.4,1.6]$. As shown in Fig.\,\ref{fig:3} and \ref{fig:4}, we find for $1.4 \lesssim n \lesssim 1.45$ that $\lambda^{(0)}$ is in the $A_{2}$ representation with a single-component order parameter $\Delta(\bk)\propto\Delta_{1}(\bk)$. In contrast, for all other $n$, we find that $\lambda^{(0)}$ is in the $E$ representation with a two-component order parameter, $\Delta(\bk)=\eta_{1}\Delta_{1}(\bk)+\eta_{2}\Delta_{2}(\bk)$. Notably, we also observe that the magnitude of $\lambda^{(0)}$, which we associate with the pairing strength, is greatly enhanced for $n\approx0.9$. This enhanced pairing strength 
leads to a significant enhancement in the critical temperature $T_{c}$ and suggest a pronounced tendency towards superconductivity at $n=0.9$. 

To understand the origin of the $E$ representation superconductivity, we note that the DOS of $\xi_{+}(\bk)$ is maximal near the saddle points $\bQ'_{j}=C^{j-1}_{3}\bkappa_{-}$ for $n\approx 1$. The amplitude of the scattering processes between these points is $\propto\chi\equiv\chi_{+-}(\bQ'_{i}+\bQ'_{j})$, which are exactly the points where $\chi_{+-}(\bq)$ is maximal. We thus construct a three-patch model for the order parameters $\boldsymbol{\Delta}=[\Delta(\bQ'_{1}),\Delta(\bQ'_{2}),\Delta(\bQ'_{3})]^{T}$ that (analogous to Eq.\,\eqref{Eq8}) is given by the eigenvalue equation $\chi(\boldsymbol{P}-\boldsymbol{1})\boldsymbol{\Delta}=\lambda\boldsymbol{\Delta}$ with $(\boldsymbol{P})_{ij}=1$ and $(\boldsymbol{1})_{ij}=\delta_{ij}$. The most negative eigenvalue $\lambda^{(0)}=-\chi$ is indeed found to be doubly degenerate with eigenvectors $(-1,1,0)^{T}$ and $(1,1,-2)^{T}$ in the $E$ representation. In contrast, the positive eigenvalue $\lambda^{(1)}=2\chi$ is in the $A$ representation with eigenvector $(1,1,1)^{T}$.

\subsection{Ginzburg-Landau analysis}
Lastly, we want to identify the superconducting state $\boldsymbol{\eta}=(\eta_{1},\eta_{2})$ realized in the $E$ representation of our model.
For that purpose, we write down the ($C_{3v}$-symmetric) Ginzburg-Landau free energy for our system,
\begin{equation}
F_{\text{GL}}=
a(T-T_{c})\,
\boldsymbol{\eta}
\cdot
\boldsymbol{\eta}^{*}
+
b_{1}
(
\boldsymbol{\eta}
\cdot
\boldsymbol{\eta}^{*}
)^{2}
+
b_{2}
|
\boldsymbol{\eta}
\cdot
\boldsymbol{\eta}
|^{2}
+
...,
\end{equation}
where $a>0$ induces the superconducting phase if $T<T_{c}$ and $b_{1}>0$ ensures thermodynamic stability. The sign of $b_{2}$ fixes the form of the superconducting state: If $b_{2}<0$, $F_{\text{GL}}$ is minimized by the \textit{nematic state} with $\boldsymbol{\eta}=(\cos\varphi,\sin\varphi)$ and $\varphi\in[0,2\pi)$. The latter spontaneously breaks the three-fold rotation symmetry of the lattice but preserves time-reversal symmetry. If $b_{2}>0$, $F_{\text{GL}}$ is minimized by the \textit{chiral state} with $\boldsymbol{\eta}=(1,\pm i)$. The chiral state breaks time-reversal symmetry spontaneously and is characterized by chiral Majorana edge states at its boundaries. 

Considering the possibility of both a chiral state or a nematic state in tWSe$_{2}$, an interesting question is how these possible pairing states could be detected in a possible future experiments. For example, an experimental signature of the Majorana edge modes in the chiral state is a quantized thermal Hall conductivity. Another possibility is to use the Polar-Kerr effect to detect the time-reversal symmetry breaking of the superconducting order parameter in the chiral state. The nematic state, on the other hand, could be distinguished by anisotropies in the upper critical field and optical responses 
\cite{bib:Venderbos2016,bib:Cao2021_2,bib:Jin2021}.

\section{Discussion and conclusion}
Having identified the pairing symmetry for a superconducting state in tWSe$_{2}$, it is now instructive to compare our theoretical considerations 
to the experimental results of Wang \textit{et al}\cite{bib:Wang2020}.
As already discussed in introduction, this experiment reported the emergence of a putative superconducting state in tWSe$_{2}$ upon doping way from a correlated insulator state at half-filling. In the experiment, two features pointed towards the possible emergence of superconductivity: 
(1) A significant reduction in resistance below approximately $3\,K$ at a twist angle of $5.1^{\circ}$, near half-filling, and under the influence of a displacement field, and (2) a flattening of the current-voltage curve below about $2\,K$ under the same conditions. It should be noted that the zero-resistance state observed in Ref.\cite{bib:Wang2020} ``was unstable to repeated thermal cycling". Further experimental study is needed to establish reproducible superconductivity in tWSe$_{2}$.   

In our theory work, we introduced a realistic model for superconductivity in tWSe$_{2}$ based on spin-valley fluctuations. We have shown that the enhancement of spin-valley fluctuations around half-filling can provide a mechanism for Cooper pairing and predicted an unconventional, two-component order parameter for superconductivity in tWSe$_{2}$, realizing either a 
chiral topological superconductor or a time-reversal invariant nematic superconductor. In future studies, it will be interesting to explore experimental signatures of the predicted two-component order parameter. 

\textit{Note added.} Very recently, a new study reported the observation of  unconventional superconductivity in tWSe$_{2}$\cite{bib:Xia2024}. 

\begin{acknowledgments}
This work was supported by a Simons Investigator Award from the Simons Foundation. LF was supported, in part, by the Air Force Office of Scientific Research (AFOSR) under award FA9550-22-1-0432.
\end{acknowledgments}

\begin{widetext}

\newpage

\onecolumngrid

\bigskip 

\begin{center}
\setcounter{page}{1}
\large{\bf Supplemental Material to `Nematic, chiral and topological superconductivity in transition metal dichalcogenides' \\}
\end{center}
\vspace{-7pt}
\begin{center}
Constantin Schrade$^{1}$ and Liang Fu$^{2}$\\
\vspace{2pt}
\it{$^{1}$ Hearne Institute of Theoretical Physics, Department of Physics \& Astronomy, Louisiana State University, Baton Rouge LA 70803, USA}
\\
\it{$^{2}$ Department of Physics, Massachusetts Institute of Technology, 77 Massachusetts Ave., Cambridge, MA 02139}
\end{center}
In the Supplemental Material, we provide details on the derivation of the interaction vertex, the self-consistency
equation, and the Ginzburg-Landau free energy. We also discuss an additional effective triangular-lattice Hubbard model.
\section{Effective pairing interaction}
In this first section of the Supplemental Material (SM), we provide details on the derivation of the effective pairing interaction from a Kohn-Luttinger renormalization of the bare repulsive interaction.
\\-
As a first step in this derivation, we write the partition function of the Hamiltonian $H$ (as given in Eq.\,(6) of the main text) in terms of a 
coherent state Feynman path integral over Grassman variables,
\begin{equation}
\begin{split}
Z&= 
\int\mathcal{D}[c^{*},c]\, e^{-S_{0}-S_{I}}
,\\
S_{0}&=
\sum_{s}
\sum_{\bk,\omega_{n}}
c^{*}_{\bk,s}(i\omega_{n})
\left[
-i\omega_{n}+\xi_{s}(\bk)
\right]
c_{\bk,s}(i\omega_{n}),
\\
S_{I}&=
\frac{U}{\beta N}\sum_{s,s'}
\sum_{\substack{\bk,\bk',\bq \\ \omega_{n},\omega'_{n},\omega_{\nu}}} 
c^{*}_{\bk+\bq,s}(i\omega_{n}+i\omega_{\nu}) c^{*}_{\bk'-\bq,s'}(i\omega'_{n}-i\omega_{\nu})c_{\bk',s'}(i\omega'_{n})c_{\bk,s}(i\omega_{n}).
 \end{split}
\end{equation}
Next, we introduce a low-energy effective action for modes in the vicinity of the Fermi surfaces \cite{bib:Shankar1994,bib:Vafek2014}, 
\begin{equation}
S^{\text{eff}}_{I}
=
\langle
S_{I}
\rangle
-
\frac{1}{2}
\left(
\langle
S^{2}_{I}
\rangle
-
\langle
S_{I}
\rangle^{2}
\right),
\end{equation}
where $\langle ... \rangle$ denotes the averaging over `high' energy modes with action $S_0$. We note that the first term in the expression for $S^{\text{eff}}_{I}$ corresponds to the bare repulsive interaction and is $\propto U$. In contrast, the second term renormalizes the bare repulsive interaction and is $\propto U^{2}$. 
\\
\\
To evaluate the renormalization of the bare repulsive interaction, we write it in the more compact form, 
\begin{equation}
S^{\text{eff}}_{I}
-
\langle
S_{I}
\rangle
=
-
\frac{1}{2}
\left[
\sum_{1,2,3,4}
\sum_{1',2',3',4'}
\langle
c^{*}_{1}
 c^{*}_{2}
 c_{3}
c_{4}
 c^{*}_{1'}
  c^{*}_{2'}
c_{3'}
  c_{4'}
 \rangle
 -
 \langle
S_{I}
\rangle^{2}
 \right]
 =
-
\frac{1}{2}
\sum_{1,2,3,4}
\sum_{1',2',3',4'}
\langle
c^{*}_{1}
 c^{*}_{2}
 c_{3}
c_{4}
 c^{*}_{1'}
  c^{*}_{2'}
c_{3'}
  c_{4'}
 \rangle_{\text{con}}. 
\end{equation}
Here, we have introduced the short-hand notation $\sum_{1,2,3,4}\equiv(U/\beta N)\sum_{s,s'}\sum_{\bk,\bk',\bq}\sum_{\omega_{n},\omega'_{n},\omega_{\nu}}$. 
Moreover, in the second equality, we have noted that the Wick contractions which correspond to disconnected Feynman diagrams in $\langle
S^{2}_{I}\rangle$ are cancelled by $ \langle
S_{I}
\rangle^{2}$. In particular, the average $\langle ... \rangle_{\text{con}}$ only involves Wick contractions that correspond to connected Feynman diagrams. 
\\
\\
Within the Kohn-Luttinger approach, the corrections to the bare repulsive interactions are given by the Wick contractions in the particle-hole channel. These Wick contractions are classified into three types, 
\begin{equation}
\begin{split}
\text{Type 1:}\qquad
&\langle
\wick{
 \c1  c^{*}_{1}
 c^{*}_{2}
 c_{3}
  \c2 c_{4}
 \c2 c^{*}_{1'}
  c^{*}_{2'}
 \c1 c_{3'}
  c_{4'}
 }
 \rangle
,
\quad\langle
\wick{
 \c1  c^{*}_{1}
 c^{*}_{2}
 c_{3}
  \c2 c_{4}
c^{*}_{1'}
  \c2  c^{*}_{2'}
 c_{3'}
 \c1  c_{4'}
 }
 \rangle
,\quad\langle
\wick{
  c^{*}_{1}
 \c1 c^{*}_{2}
  \c2 c_{3}
 c_{4}
  \c2 c^{*}_{1'}
  c^{*}_{2'}
 \c1 c_{3'}
  c_{4'}
 }
 \rangle
,\quad\langle
\wick{
  c^{*}_{1}
 \c1 c^{*}_{2}
  \c2 c_{3}
 c_{4}
c^{*}_{1'}
  \c2   c^{*}_{2'}
c_{3'}
 \c1   c_{4'}
 }
 \rangle
\\
&\langle
\wick{
 \c1  c^{*}_{1}
 c^{*}_{2}
   \c2 c_{3}
c_{4}
 \c2 c^{*}_{1'}
  c^{*}_{2'}
c_{3'}
 \c1   c_{4'}
 }
 \rangle
,
\quad\langle
\wick{
 c^{*}_{1}
 \c1  c^{*}_{2}
 c_{3}
  \c2 c_{4}
 \c2 c^{*}_{1'}
 c^{*}_{2'}
 c_{3'}
  \c1  c_{4'}
 }
 \rangle
,\quad\langle
\wick{
 \c1  c^{*}_{1}
 c^{*}_{2}
  \c2 c_{3}
 c_{4}
 c^{*}_{1'}
  \c2  c^{*}_{2'}
 \c1 c_{3'}
  c_{4'}
 }
 \rangle
,\quad\langle
\wick{
  c^{*}_{1}
  \c1 c^{*}_{2}
c_{3}
 \c2  c_{4}
c^{*}_{1'}
  \c2   c^{*}_{2'}
 \c1 c_{3'}
   c_{4'}
 }
 \rangle
 \\
 \\
\text{Type 2:}\qquad
&\langle
\wick{
 c^{*}_{1}
 \c1  c^{*}_{2}
  \c2 c_{3}
 c_{4}
 \c2 c^{*}_{1'}
  c^{*}_{2'}
 c_{3'}
 \c1  c_{4'}
 }
 \rangle
,
\quad\langle
\wick{
c^{*}_{1}
 \c1   c^{*}_{2}
  \c2  c_{3}
c_{4}
c^{*}_{1'}
  \c2  c^{*}_{2'}
  \c1  c_{3'}
c_{4'}
 }
 \rangle
,\quad\langle
\wick{
 \c1  c^{*}_{1}
 c^{*}_{2}
c_{3}
  \c2  c_{4}
c^{*}_{1'}
   \c2  c^{*}_{2'}
 \c1  c_{3'}
 c_{4'}
 }
 \rangle
,\quad\langle
\wick{
 \c1   c^{*}_{1}
c^{*}_{2}
c_{3}
  \c2  c_{4}
  \c2 c^{*}_{1'}
  c^{*}_{2'}
 c_{3'}
 \c1  c_{4'}
 }
 \rangle
 \\
 \\
\text{Type 3:}\qquad
&\langle
\wick{
 c^{*}_{1}
 \c1  c^{*}_{2}
 c_{3}
  \c2 c_{4}
 \c2 c^{*}_{1'}
  c^{*}_{2'}
 \c1 c_{3'}
  c_{4'}
 }
 \rangle
,
\quad\langle
\wick{
 \c1  c^{*}_{1}
 c^{*}_{2}
 \c2  c_{3}
 c_{4}
c^{*}_{1'}
  \c2  c^{*}_{2'}
 c_{3'}
 \c1  c_{4'}
 }
 \rangle
,\quad\langle
\wick{
 \c1  c^{*}_{1}
 c^{*}_{2}
  \c2 c_{3}
 c_{4}
  \c2  c^{*}_{1'}
 c^{*}_{2'}
 \c1  c_{3'}
 c_{4'}
 }
 \rangle
,\quad\langle
\wick{
c^{*}_{1}
 \c1   c^{*}_{2}
c_{3}
  \c2  c_{4}
c^{*}_{1'}
  \c2   c^{*}_{2'}
 c_{3'}
  \c1 c_{4'}
 }
 \rangle
  \end{split}
\end{equation}
For each of these three types of Wick contractions, we show example Feynman diagrams in Fig.\,5. We now proceed with the evaluation of the different types of Wick contractions:
\subsection{Wick contractions of type 1 and 2}
We begin by evaluation the Wick contractions of type 1 and, for that purpose, consider the following example,
\begin{align}
&\ \ \ -\frac{1}{2}\sum_{1,2,3,4}
\sum_{1',2',3',4'}
\langle
\wick{
 \c1  c^{*}_{1}
 c^{*}_{2}
 c_{3}
  \c2 c_{4}
 \c2 c^{*}_{1'}
  c^{*}_{2'}
 \c1 c_{3'}
  c_{4'}
 }
 \rangle
 \\
 &=
- \frac{1}{2(\beta N)^{2}}
\sum_{\substack{\bk,\bk',\bq \\ \omega_{n},\omega'_{n},\omega_{\nu}}} 
G_{\ua}(\bq,i\omega_{\nu})G_{\ua}(\bq+\bk-\bk',i\omega_{\nu}
+
i\omega_{n}
+
i\omega'_{n}
)
\langle
c^{*}_{\bk,\ua}(i\omega_{n})
c^{*}_{-\bk,\da}(-i\omega_{n})
c_{-\bk',\da}(-i\omega'_{n})
c_{\bk',\ua}(i\omega'_{n})
\rangle\nonumber
 \\
 &
 \quad - \frac{1}{2(\beta N)^{2}}
\sum_{\substack{\bk,\bk',\bq \\ \omega_{n},\omega'_{n},\omega_{\nu}}} 
G_{\da}(\bq,i\omega_{\nu})G_{\da}(\bq+\bk-\bk',i\omega_{\nu}
+
i\omega_{n}
+
i\omega'_{n}
)
\langle
c^{*}_{\bk,\ua}(i\omega_{n})
c^{*}_{-\bk,\da}(-i\omega_{n})
c_{-\bk',\da}(-i\omega'_{n})
c_{\bk',\ua}(i\omega'_{n})
\rangle\nonumber
\end{align}
In the second line of the example, we have defined the Green's function $G_{s}(\bk,i\omega_n)=\langle c^{\dag}_{\bk,s}(i\omega_{n})c_{\bk,s}(i\omega_{n})\rangle$ and only
kept the terms that lead to a pairing interaction. The remaining Wick contractions  of type 1 give the same result and, therefore, we do not display them here.
\\
\\
Next, we want to compare this finding with a particular example for a Wick contraction of type 2, 
\begin{align}
& -\frac{1}{2}\sum_{1,2,3,4}
\sum_{1',2',3',4'}
\langle
\wick{
 c^{*}_{1}
 \c1  c^{*}_{2}
  \c2 c_{3}
 c_{4}
 \c2 c^{*}_{1'}
  c^{*}_{2'}
 c_{3'}
 \c1  c_{4'}
 }
 \rangle
\\
 &\equiv
 \frac{1}{(\beta N)^{2}}
\sum_{\substack{\bk,\bk',\bq \\ \omega_{n},\omega'_{n},\omega_{\nu}}} 
G_{\ua}(\bq,i\omega_{\nu})G_{\ua}(\bq+\bk-\bk',i\omega_{\nu}
+
i\omega_{n}
+
i\omega'_{n}
)
\langle
c^{*}_{\bk,\ua}(i\omega_{n})
c^{*}_{-\bk,\da}(-i\omega_{n})
c_{-\bk',\da}(-i\omega'_{n})
c_{\bk',\ua}(i\omega'_{n})
\rangle\nonumber
 \\
 &
+ \frac{1}{(\beta N)^{2}}
\sum_{\substack{\bk,\bk',\bq \\ \omega_{n},\omega'_{n},\omega_{\nu}}} 
G_{\da}(\bq,i\omega_{\nu})G_{\da}(\bq+\bk-\bk',i\omega_{\nu}
+
i\omega_{n}
+
i\omega'_{n}
)
\langle
c^{*}_{\bk,\ua}(i\omega_{n})
c^{*}_{-\bk,\da}(-i\omega_{n})
c_{-\bk',\da}(-i\omega'_{n})
c_{\bk',\ua}(i\omega'_{n})
\rangle\nonumber
\end{align}
Here, we observe that this finding differs from the previous one by a minus sign and a prefactor of 2. Since the remaining 
Wick contractions of type 2 all give the same result as this example, we find that the type 1 and type 2 contributions cancel each other and, therefore, do not 
contribute to the effective action.
 \begin{figure}[!b] \centering
\includegraphics[width=\linewidth] {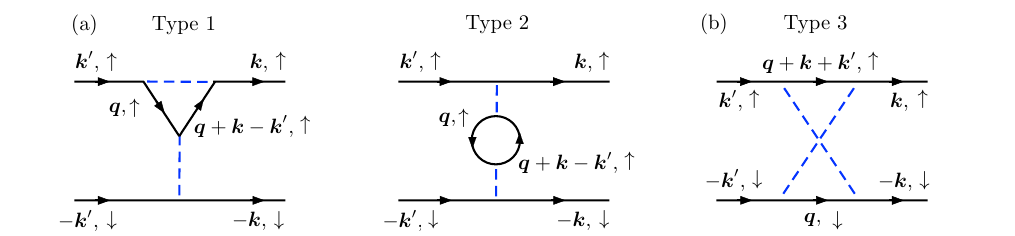}
\caption{(Color online)
(a) Feynman diagrams corresponding to Wick contractions of type 1 and type 2 within the Kohn-Luttinger renormalization of the bare repulsive interaction. The closed fermion loop in the Feynman diagram of type 2 produces an additional minus sign and, therefore, the Feynman diagrams of type 1 and 2 cancel each other. (b) Feynman diagram corresponding to Wick contractions of type 3, which yield a finite contribution to the effective pairing interaction.
}\label{fig:5}
\end{figure}

\subsection{Wick contractions of type 3}
For illustrating the evaluation of a Wick contraction of type 3, we again pick a specific example,
\begin{align}
&\ \ \ -\frac{1}{2}\sum_{1,2,3,4}
\sum_{1',2',3',4'}
\langle
\wick{
 c^{*}_{1}
 \c1  c^{*}_{2}
 c_{3}
  \c2 c_{4}
 \c2 c^{*}_{1'}
  c^{*}_{2'}
 \c1 c_{3'}
  c_{4'}
 }
 \rangle
 \\
&\equiv
- \frac{1}{(\beta N)^{2}}
\sum_{\substack{\bk,\bk',\bq \\ \omega_{n},\omega'_{n},\omega_{\nu}}} 
G_{\da}(\bq,i\omega_{\nu})G_{\ua}(\bq+\bk+\bk',i\omega_{\nu}
+
i\omega_{n}
+
i\omega'_{n}
)
\langle
c^{*}_{\bk,\ua}(i\omega_{n})
c^{*}_{-\bk,\da}(-i\omega_{n})
c_{-\bk',\da}(-i\omega'_{n})
c_{\bk',\ua}(i\omega'_{n})
\rangle\nonumber
\rangle.
\end{align}
Similar to the previous subsection, we find that all type 3 Wick contractions give the same contribution as this example. 
Hence, we find that the amplitude of the effective pairing interaction reads, 
\begin{equation}
\begin{split}
\Gamma_{}(\bk,\bk')=U+U^{2}\chi_{+-}(\bk+\bk')
\quad \text{with}\quad \chi_{+-}(\bq)&\equiv \chi_{+-}(\bq,i\omega=0)
\\
&=-\frac{1}{\beta N}\sum_{\bk}
G_{\da}(\bq,i\omega=0)G_{\ua}(\bq+\bk,i\omega=0)
\\
&=
-
\frac{1}{N}
\sum_{\bk}
\frac{
f(\xi_{\da}(\bk)-\mu)-f(\xi_{\ua}(\bq+\bk)-\mu)
}{\xi_{\da}(\bk)-\xi_{\ua}(\bq+\bk)}.
\end{split}
\end{equation}
Here, we have set the Matsubara frequencies at the external legs of the Feynman diagrams to zero and omitted numerical prefactors in $\chi_{+-}$ as they will not be relevant for the weak coupling study of superconductivity. 
Furthermore, we have introduced the Fermi distribution $f$ and carried out the summation over the Matsubara frequencies by using
$G_{s}(\bk,i\omega_n)=1/(i\omega_{n}-\xi_{s}(\bk)+\mu)$.  
This concludes our derivation of the effective pairing interaction.

\section{Mean-field theory and self-consistency equation}
In this second section of the SM, we provide details on the derivation of the pairing Hamiltonian within mean-field theory and determine the self-consistency equation for the superconducting order parameter. 
\\

In a first step, we consider the effective pairing interaction in the Hamiltonian form, 
\begin{equation}
H_{I}^{\text{eff}}
=
\frac{1}{N}\sum_{\bk,\bk'}
\Gamma_{}(\bk,\bk')\,
c^{\dag}_{\bk, \ua}c^{\dag}_{-\bk, \da}c_{-\bk', \da}c_{\bk', \ua}
\quad
\text{with}
\quad
\Gamma_{}(\bk,\bk')
=
U
+
U^{2}
\chi_{+-}(\bk+\bk'),
\end{equation}
and perform a mean-field decoupling,
\begin{equation}
\begin{split}
H_{I}^{\text{eff}}
\rightarrow H_{\Delta}
&=
\frac{1}{2N}\sum_{\bk,\bk'}
\Gamma_{}(\bk,\bk')\,
\left[
\langle c^{\dag}_{\bk, \ua}c^{\dag}_{-\bk, \da}\rangle c_{-\bk', \da}c_{\bk', \ua}
+
c^{\dag}_{\bk, \ua}c^{\dag}_{-\bk, \da} \langle c_{-\bk', \da}c_{\bk', \ua}\rangle
\right]
\\
&=
-
\sum_{\bk}
\Delta(\bk)\,
c_{-\bk, \da}c_{\bk, \ua}
-
\sum_{\bk}
\Delta(\bk)^{*}\,
c^{\dag}_{\bk, \ua}c^{\dag}_{-\bk, \da},
\end{split}
\end{equation}
where we have defined the superconducting order parameter as,
\begin{equation}
\begin{split}
\Delta(\bk)
&=
-\frac{1}{2N}
\sum_{\bk'}
\Gamma_{}(\bk,\bk')\,
\langle
c^{\dag}_{\bk', \ua}c^{\dag}_{-\bk', \da}
\rangle.
\end{split}
\end{equation}

In a second step, we write the pairing Hamiltonian $H_{\Delta}$ and the normal state Hamiltonian $H_{0}$ as a Bogoliubov-de Gennes Hamiltonian matrix,
\begin{equation}
\begin{split}
H&=
H_{0}+H_{\Delta}
\\
&=
\sum_{\bk} \xi_{\ua}(\bk)\,c^{\dag}_{\bk,\ua} c_{\bk,\ua} 
+
 \xi_{\da}(\bk)\,c^{\dag}_{\bk,\da} c_{\bk,\da} 
 -
 \Delta(\bk)\,
c_{-\bk, \da}c_{\bk, \ua}
-\Delta(\bk)^{*}\,
c^{\dag}_{\bk, \ua}c^{\dag}_{-\bk, \da}
\\
&=
\sum_{\bk} \xi_{\ua}(\bk)\left[ c^{\dag}_{\bk,\ua} c_{\bk,\ua} 
+
c^{\dag}_{-\bk,\da} c_{-\bk,\da} \right]
 -
 \Delta(\bk)\,
c_{-\bk, \da}c_{\bk, \ua}
-\Delta(\bk)^{*}\,
c^{\dag}_{\bk, \ua}c^{\dag}_{-\bk, \da}
\\
&=
\sum_{\bk}
\begin{pmatrix}
c^{\dag}_{\bk,\ua} &
c_{-\bk,\da}
\end{pmatrix}
\begin{pmatrix}
\xi_{\ua}(\bk) & -\Delta(\bk)^{*}\\
-\Delta(\bk) & -\xi_{\ua}(\bk)
\end{pmatrix}
\begin{pmatrix}
c_{\bk,\ua} \\
c^{\dag}_{-\bk,\da}
\end{pmatrix}
+
E_{0}
\end{split}
\end{equation}
where we have used that $\xi_{\da}(-\bk)=\xi_{\ua}(\bk)$ and defined the energy offset $E_{0}$. We now introduce the Bogoliubov transformations,
\begin{equation}
\begin{split}
\begin{pmatrix}
c_{\bk,\ua} \\
c^{\dag}_{-\bk,\da}
\end{pmatrix}
&=
\begin{pmatrix}
u^{*}_{\bk}& v_{\bk} \\
-v_{\bk}& u_{\bk}
\end{pmatrix}
\begin{pmatrix}
\gamma_{\bk,1} \\
\gamma^{\dag}_{-\bk,2}
\end{pmatrix}
\ \
\text{with}
\ \
u_{\bk}=
\frac{\Delta(\bk)}{|\Delta(\bk)|}
\sqrt{\frac{1}{2}\left(1+\frac{\xi_{\ua}(\bk)}{E_{}(\bk)}\right)},
\quad
v_{\bk}=
\sqrt{\frac{1}{2}\left(1-\frac{\xi_{\ua}(\bk)}{E_{}(\bk)}\right)}, 
\end{split}
\end{equation}
and apply it to the Bogoliubov-de Gennes Hamiltonian matrix. This yields the diagonalized Hamiltonian, 
\begin{equation}
\begin{split}
H&=
\sum_{\bk}
\begin{pmatrix}
\gamma^{\dag}_{\bk,1} &
\gamma_{-\bk,2}
\end{pmatrix}
\begin{pmatrix}
E_{}(\bk) & 0\\
0 & -E_{}(\bk)
\end{pmatrix}
\begin{pmatrix}
\gamma_{\bk,1} \\
\gamma^{\dag}_{-\bk,2},
\end{pmatrix}
=
\sum_{\bk}
E(\bk) \left[ 
\gamma^{\dag,1}_{\bk}\gamma_{\bk,1}
+
\gamma^{\dag}_{-\bk,2}
\gamma_{-\bk,2}
\right]
+
E'_{0},
\end{split}
\end{equation}
where we have defined the quasiparticle energy $E(\bk) 
=
\sqrt{\xi_{\ua}(\bk)^{2}+|\Delta(\bk)|^{2}}$
and
another energy offset $E'_{0}$.
\\

In a third step, we evaluate the expectation value,
\begin{equation}
\begin{split}
\langle c^{\dag}_{\bk,\ua}c^{\dag}_{-\bk,\da}\rangle
&=
\frac{1}{2}
\left[
-
u_{\bk}v_{\bk}
\langle \gamma^{\dag}_{\bk,1}\gamma_{\bk,1}\rangle
+
v_{\bk}u_{\bk}
\langle \gamma_{-\bk,2}\gamma^{\dag}_{-\bk,2}\rangle
\right]
\\
&=
\frac{1}{2}
u_{\bk}v_{\bk}
\left[
1
-
\langle \gamma^{\dag}_{\bk,1}\gamma_{\bk,1}\rangle
-
\langle\gamma^{\dag}_{-\bk,2} \gamma_{-\bk,2}\rangle
\right]
\\
&=
\frac{1}{2}
u_{\bk}v_{\bk}
\left[
1-2f(E(\bk))
\right]
\\
&=
\frac{1}{2}
u_{\bk}v_{\bk}
\tanh\left(
\frac{E(\bk)}{2k_{B}T}
\right)
\\
&=
\frac{1}{4}
\frac{\Delta(\bk)}{E(\bk)}
\tanh\left(
\frac{E(\bk)}{2k_{B}T}
\right)
. 
\end{split}
\end{equation}
Here, $f$ denotes the Fermi-Dirac distribution. We now reinsert the expression of the expectation value in the definition of the superconducting order parameter. Under the assumption that the superconducting order parameter is sufficiently small, we find the self-consistency equation,
\begin{equation}
\begin{split}
\Delta_{}(\bk)
&=
-\frac{1}{8N}
\sum_{\bk'}
\Gamma_{}(\bk,\bk')\,
\frac{\Delta_{}(\bk')}{\xi_{\ua}(\bk')}
\tanh\left(
\frac{\xi_{\ua}(\bk')}{2k_{B}T}
\right).
\end{split}
\end{equation}
\\

In a final step, we rewrite this self-consistency equation as an integral eigenvalue equation,
\begin{equation}
\int_{\text{FS}_{\ua}} \frac{\mathrm{d}\hat\bk'}{v_{\ua}(\hat\bk')}
\Gamma_{}(\hat\bk,\hat\bk')
\Delta_{}(\hat\bk')
=
\lambda
\Delta_{}(\hat\bk),
\end{equation}
which determines the superconducting order parameter on the $+K$ Fermi surface, $\xi_{\ua}(\bk)-\mu=0$, with $v_{\ua}(\bk)=||\partial \xi_{\ua}(\bk)/\partial\bk||$ denoting the Fermi velocity. We note that the superconducting order parameters at the $+K$ and $-K$ Fermi surfaces are given by $\Delta(\bk)$ and $\Delta'(\bk)=-\Delta(-\bk)$, respectively. This concludes our derivation of the self-consistency equation of the superconducting order parameter.

\section{Numerical evaluation of the self-consistency equation}
In this third section of the SM, we give details on the numerical evaluation of the self-consistency equation. 
\\
\\
To start, we write the line integral in the self-consistency equation as an area integral over a small neighborhood 
of the $+K$ Fermi surface given by the level set $\xi_{\ua}(\bk)-\mu=\pm\Omega$, 
\begin{equation}
\int_{\text{FS}_{\ua}} \frac{\mathrm{d}\hat\bk'}{v_{\ua}(\hat\bk')}
\Gamma_{}(\hat\bk,\hat\bk')
\Delta_{}(\hat\bk')
\approx
\frac{1}{2\Omega}
\int_{|\xi_{\ua}(\bk')-\mu|<\Omega}
\mathrm{d}\bk'\,
\Gamma_{}(\hat\bk,\hat\bk')
\Delta_{}(\hat\bk')
\approx
\frac{1}{2\Omega}
\sum_{\substack{\bk' \\ |\xi_{\ua}(\bk')-\mu|<\Omega}} 
\delta\bk'\,
\Gamma_{}(\hat\bk,\hat\bk')
\Delta_{}(\hat\bk').
\end{equation}
Here, we have noted that the width of the area defined by the level set is roughly $2\Omega|\nabla_{\bk}\xi_{\ua}(\bk)|^{-1}=2\Omega\, v_{\ua}(\hat\bk')^{-1}$ and divided the integrand by this factor. Subsequently, we have written the area integral in terms of a Riemann sum. 
\\
\\
There are now two alternative approaches for numerically solving the eigenvalue equation:
\\

A first approach is to fix a set of vectors $\{\bk_{i} \}$ in momentum space within the window $|\xi_{+}(\bk)-\mu|<\Omega$ and write the Riemann sum 
as a matrix $\boldsymbol{A}$ with entries  $[\boldsymbol{A}]_{ij}=\Gamma(\bk_{i},\bk'_{j})\,\delta\bk'_{j}/2\Omega$. Diagonalization of this matrix yields the desired eigenvalues $\lambda$
and eigenvectors $\Delta(\bk)$. However, it is important to note that within this approach the resulting eigenvectors are not automatically classified in terms of the irreducible representations of C$_{3v}$. 
\\

A second approach is to fix an irreducible representation $D$ of C$_{3v}$ and expand the superconducting order parameter in terms of basis functions
that are orthogonal with respect to the Fermi surface inner product, 
\begin{equation}
\Delta^{D}(\bk)=\sum_{i} c^{D}_{i} \psi^{D}_{i}(\bk)
\quad
\text{with}
\quad
\left(\frac{1}{2\Omega}\right)^{2}
\sum_{\substack{\bk,\bk' \\ |\xi_{\ua}(\bk)-\mu|<\Omega\\ |\xi_{\ua}(\bk')-\mu|<\Omega}} 
\psi_{i}^{D}(\bk)^{*}\psi_{j}^{D}(\bk')\,\delta\bk\,\delta\bk'
=
\delta_{ij}.
\end{equation}
This expansion allows us to write the self-consistency equation in the matrix form, 
\begin{equation}
\boldsymbol{A}^{D}\boldsymbol{c}^{D}=\lambda^{D}\boldsymbol{c}^{D}
\quad
\text{with}
\quad
[\boldsymbol{A}^{D}]_{ij}=
\left(\frac{1}{2\Omega}\right)^{2}
\sum_{\substack{\bk,\bk' \\ |\xi_{\ua}(\bk)-\mu|<\Omega\\ |\xi_{\ua}(\bk')-\mu|<\Omega}} 
\psi_{i}^{D}(\bk)^{*}\Gamma(\hat\bk,\hat\bk')\psi_{j}^{D}(\bk')\,\delta\bk\,\delta\bk',
\end{equation}
so that we can solve for $\boldsymbol{c}^{D}=(c_{1},c_{2},...)^{T}$ to obtain the eigenvalues $\lambda^{D}$ and the eigenvectors $\Delta^{D}(\bk)$. 
\\
\\
Finally, we outline a two-step protocol \cite{bib:Platt2013} to generate the orthogonal basis functions for a $d$-dimensional representation $D$:
As a first step, we fix a set of linear-independent vectors $\{\bdelta_{n,i}\}_{i=1,..,d}$ that connect a site on the superlattice with an $n$-th nearest neighbor site. This set of vectors allows us to define the lattice harmonics, 
\begin{equation}
\tilde\psi^{D}_{n,i}(\bk)=\sum_{G}\text{char}_{D}([G])\, e^{i\bk\cdot G\bdelta_{n,i}},
\end{equation}
where the summation is over all the group elements $G$ of $C_{3v}$ and $\text{char}_{D}([G])$ denotes the character of the conjugacy class $[G]$
in the representation $D$. In a second step, we use a Gram-Schmidt procedure to orthogonalize the lattice harmonics, leading to a desired set of orthogonal basis functions $\psi_{i}^{D}(\bk)$. This concludes our discussion on the numerical evaluation of the self-consistency equation.

\section{Ginzburg-Landau free energy}
In this fourth section of the Supplemental Material, we present details on the analysis of the phenomenological Ginzburg-Landau free energy, which allows to us to identify the candidate pairing states that is realized in the $E$-representation of our model for superconductivity in tWSe$_{2}$.
\\

In a first step, we introduce a general linear combination of the superconducting order parameter, 
\begin{equation}
\Delta(\bk)=\eta_{1}\Delta_{1}(\bk)+\eta_{2}\Delta_{2}(\bk),
\end{equation}
where $\{\Delta_{1}(\bk),\Delta_{2}(\bk)\}$ are two linear-independent basis functions of the $E$-representation and $\{\eta_{1},\eta_{2}\}$ are complex expansion coefficients.
\\

In a second step, we expand the Ginzburg-Landau free energy in terms of $\boldsymbol{\eta}=(\eta_{1},\eta_{2})$,
\begin{equation}
F_{\text{GL}} = a\,
\boldsymbol{\eta}
\cdot
\boldsymbol{\eta}^{*}
+
b_{1}
(
\boldsymbol{\eta}
\cdot
\boldsymbol{\eta}^{*}
)^{2}
+
b_{2}
|
\boldsymbol{\eta}
\cdot
\boldsymbol{\eta}
|^{2}.
\end{equation}
Here, we have required that the Ginzburg-Landau free energy is (1) a real function, (2) invariant under all symmetry operations of $C_{3v}$, (3) an even function in $\eta_{1}$ and $\eta_{2}$. Moreover, we have denoted the Ginzburg-Landau expansion coefficients by $\{a,b_{1},b_{2}\}$. We point out that $a>0$ induces the superconducting phase if $T<T_{c}$. Also, we note that $|\boldsymbol{\eta}^{*}\times\boldsymbol{\eta}|$ is not an independent quartic term because $|
\boldsymbol{\eta}
\cdot
\boldsymbol{\eta}
|^{2}=(
\boldsymbol{\eta}
\cdot
\boldsymbol{\eta}^{*}
)^{2}-|\boldsymbol{\eta}^{*}\times\boldsymbol{\eta}|$.\\

In a third step, we perform a variation of the Ginzburg-Landau free energy with respect to $\boldsymbol{\eta}^{*}$. This yields the requirement, 
\begin{equation}
\begin{split}
\frac{\delta F_{\text{GL}}}{\delta \boldsymbol{\eta}^{*}}=
a\, \boldsymbol{\eta} 
+
2b_{1}
\left(
\boldsymbol{\eta}\cdot\boldsymbol{\eta}^{*}
\right)\boldsymbol{\eta}
+
2b_{2}
\left(
\boldsymbol{\eta}\cdot\boldsymbol{\eta}
\right)\boldsymbol{\eta}^{*}
\stackrel{!}{=}0.
\end{split}
\end{equation}
To ensure thermodynamic stability, we further require that $b_{1}\stackrel{!}{>}0$. 
\\

In a fourth step, we note that if $b_{2}>0$, the superconducting order parameter needs to satisfy $\boldsymbol{\eta}\cdot\boldsymbol{\eta}=0$ for minimizing the Ginzburg-Landau free energy. From this constraint, we see that the superconducting order parameter needs to take on the chiral form,
\begin{equation}
\Delta(\bk)\propto \Delta_{1}(\bk)\pm i\Delta_{2}(\bk). 
\end{equation}
\\

In a final step, we note that if $b_{2}<0$, the superconducting order parameter needs to maximize $\boldsymbol{\eta}\cdot\boldsymbol{\eta}$ for minimizing the Ginzburg-Landau free energy. This condition implies that the superconducting order parameter needs to take on the nematic form,
\begin{equation}
\Delta(\bk)\propto \cos\varphi\,\Delta_{1}(\bk)+ \sin\varphi\,\Delta_{2}(\bk),
\end{equation}
where $\varphi$ is an arbitrary phase. This concludes our derivation of the form of the superconducting order parameter.

 \begin{figure}[!b] \centering
  \centering
  \begin{minipage}[b]{0.47\textwidth}
    \includegraphics[width=\textwidth]{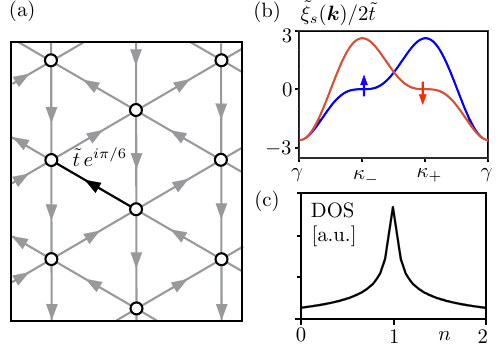}
    \caption{(Color online)
(a) Triangular lattice with directed bonds. Electrons of the $+K$ valley, pick up a tunneling phase $e^{i\pi/6}$ when hopping along the bond direction.
Electrons of the $-K$ valley pick up the opposite phase. 
(b) Dispersion $\tilde\xi_{\ua}(\bk)/2\tilde{t}$ along high-symmetry lines of the moir\'{e} Brillouin zone for the effective `$\pi/6$'-model.  
(c) Density of states of the band structure of the effective `$\pi/6$'-model. 
}\label{fig:2sm}
  \end{minipage}
  \hfill
  \begin{minipage}[b]{0.47\textwidth}
    \includegraphics[width=\textwidth]{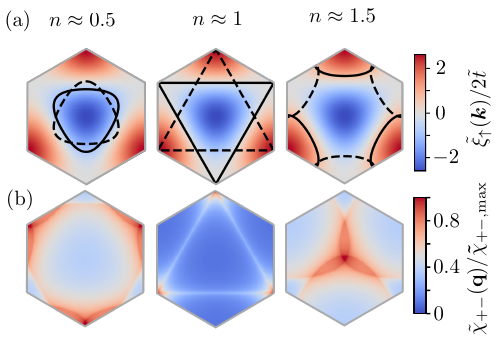}
    \caption{(Color online)
(a) Dispersion $\tilde\xi_{\ua}(\bq)$ of the top-most band at the $+K$ valleys. The  $+K$ ($-K$) valley Fermi surfaces are shown with solid (dashed) black lines. (b) Transversal spin-valley susceptibility $\tilde\chi_{+-}(\bq)$ at $1/\beta=100\,$meV. 
(c) Longitudinal spin-valley susceptibility $\tilde\chi_{zz}(\bq)$ at $1/\beta=100\,$meV. 
}\label{fig:2sm}
  \end{minipage}
\end{figure}

\section{Effective `$\pi/6$'-model}
In this seventh section of the Supplemental Material, we introduce and discuss an effective model that captures some of the features of the full continuum model of tWSe$_{2}$ for the parameter choices given in the main text. 

\subsection{Hamiltonian}
As a starting point, we introduce the Hamiltonian of the effective model,
\begin{equation}
\tilde{H}=\sum_{\bk,s}[\tilde\xi_{s}(\bk)-\mu] c^{\dag}_{\bk,s}c_{\bk,s}+\frac{U}{N}
\sum_{\substack{\bk,\bp,\bq \\ s,s'}} 
c^{\dag}_{\bq,s}c^{\dag}_{\bp+\bk-\bq,s'}c_{\bp,s'}c_{\bk,s}
\end{equation}
In this expression, we have defined the dispersion relations, which describe (as shown in Fig.\,6(a)) electron hopping on a triangular lattice, 
\begin{equation}
\tilde\xi_{s}(\bk)
=
-
2\tilde{t}
\sum_{i=1,2,3}
\cos\left(
\frac{\pi}{6}
\right)
\cos(\bk\cdot\bdelta_{i})
+
s
\sin\left(
\frac{\pi}{6}
\right)
\sin(\bk\cdot\bdelta_{i}),
\end{equation}
Here, the lattice vectors are given by $\bdelta_{1}=a_{0}/|\theta|(0,1)^{T}$, $\bdelta_{2}=a_{0}/|\theta|(\sqrt{3}/2,-1/2)^{T}$, and $\bdelta_{3}=a_{0}/|\theta|(-\sqrt{3}/2,-1/2)^{T}$. Moreover, the tunneling amplitude is denoted by $\tilde{t}$. As shown in Fig.\,6(b), the dispersion relation $\tilde\xi_{s}(\bk)$ is qualitatively similar to the dispersion relation $\xi_{s}(\bk)$ obtained from the full continuum model. In particular, the dispersion relation shows a saddle points at $\kappa_{\pm}$ that lead to a vHs at half-filling, as depicted in Fig.\,6(c).

\subsection{Comparing the effective model to the continuum model}
Next, want to compare the effective model to the continuum model of . For that purpose

we have computed the valley susceptibility from the expression,
\begin{equation}
\tilde\chi_{\alpha\beta}(\bq)
=
-
\frac{1}{\beta N}
\sum_{\bk,i\omega}
\text{Tr}\left[\sigma_{\alpha}\tilde{G}(\bk,i\omega)
\sigma_{\beta}\tilde{G}(\bk+\bq,i\omega)\right]
\quad
\text{with}
\quad
\tilde{G}_{s_{1}s_{2}}(\bk,i\nu)
=
\delta_{s_{1}s_{2}}
[
i\nu-\tilde\xi_{s_{1}}(\bk)
+\mu
]^{-1}.
\end{equation}
The results of the computation are shown in Fig.\,7. Most notably, we have found that the transversal susceptibility $\tilde\chi_{+-}(\bq)$ has a qualitatively similar structure as in the full continuum model. In particular, at half-filling, the transversal susceptibility shows the same characteristics maxima at the wavevectors $\bQ_{j}=R_{2\pi(j-1)/3}\bkappa_{+}$.
\\
\\
Finally, we compare the possible superconducting orders arising from inter-valley fluctuations in the `$\pi/6$'-model to the results of the full continuum model. As shown in Fig.\,8 and 9, the most negative eigenvalue $\tilde\lambda^{(0)}$ of the Cooper channel interaction vertex $\tilde\Gamma(\hat\bk,\hat\bk')=U+U^{2}\tilde\chi_{+-}(\hat\bk+\hat\bk')$ is in the $E$-representation of the $C_{3v}$ point group at half-filling, which agrees with the continuum model findings. However, we find that the leading superconducting order in the `$\pi/6$'-model and the full continuum model differ at quarter filling. Specifically, we find that the leading superconducting instability at quarter-filling is not in the $A_{2}$ representation as in the full continuum model, but rather in the $E$-representation. 

 \begin{figure}[!t] \centering
  \centering
  \begin{minipage}[b]{0.47\textwidth}
    \includegraphics[width=\textwidth]{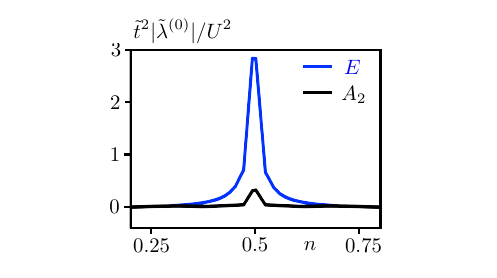}
    \caption{(Color online)
    Pairing strengths of the effective `$\pi/6$'-model given by the most negative eigenvalue $|\tilde\lambda^{(0)}|$ of the the self-consistency 
equation (within a fixed representation) as a function of the filling $n$. For fillings $0.27 \lesssim n \lesssim 0.29$ and $0.71 \lesssim n \lesssim 0.73$,
we find that the most negative eigenvalue is in the one-dimensional $A_{2}$ representation of the $C_{3v}$ point group. At all other fillings, we find that the most negative eigenvalue is in the two-dimensional $E$ representation of $C_{3v}$.
}\label{fig:2sm}
  \end{minipage}
  \hfill
  \begin{minipage}[b]{0.47\textwidth}
    \includegraphics[width=\textwidth]{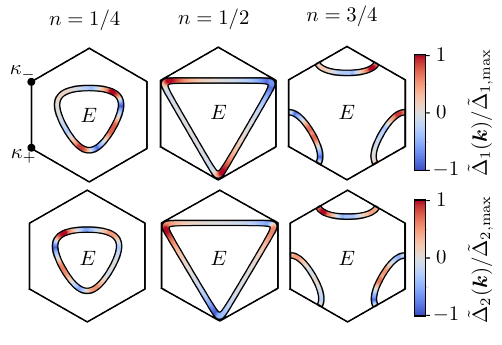}
    \caption{(Color online)
Gap functions of the leading superconducting states of the effective `$\pi/6$'-model (normalized by their maximum value).  At $n\approx1/4,\,1/2\,3/4$, the most negative eigenvalue is two-fold degenerate and the gap basis functions $\tilde\Delta_{1,2}(\bk)$ transform in the $2d$ $E$-representation of the $C_{3v}$ point group. 
}\label{fig:2sm}
  \end{minipage}
\end{figure}

{}

\end{widetext}

\end{document}